\documentclass[pdflatex,sn-mathphys]{sn-jnl}

\jyear{2023}%

\newcommand{\myparagraph}[1]{\textbf{{#1}.}}

\usepackage{amsmath}
\usepackage{bm}
\usepackage{pdfpages}
\usepackage{blindtext}

\DeclareMathOperator{\corr}{\rho}

\newcommand{\figref}{\figurename~\ref}

\usepackage{hyperref}
\hypersetup{bookmarksdepth=subsubsection}

\begin{document}

\title[Spatial gradient consistency for unsupervised demosaicking]{Spatial gradient consistency for unsupervised learning of hyperspectral demosaicking: Application to surgical imaging}

\author*[1]{\fnm{Peichao} \sur{Li}}
\email{peichao.2.li@kcl.ac.uk}

\author[1]{\fnm{Muhammad} \sur{Asad}}

\author[1]{\fnm{Conor} \sur{Horgan}}

\author[1,2]{\fnm{Oscar} \sur{MacCormac}}

\author[1,2]{\fnm{Jonathan} \sur{Shapey}}

\author[1]{\fnm{Tom} \sur{Vercauteren}}

\affil[1]{
\orgdiv{School of Biomedical Engineering \& Imaging Sciences}, 
\orgname{King's College London}, \orgaddress{\city{London}, \country{UK}}}

\affil[2]{
\orgdiv{Department of Neurosurgery},
\orgname{King's College Hospital NHS Foundation Trust}, \orgaddress{\city{London}, \country{UK}}}

\abstract{
\textbf{Purpose:}
Hyperspectral imaging has the potential to improve intraoperative decision making if tissue characterisation is performed in real-time and with high-resolution.
Hyperspectral snapshot mosaic sensors offer a promising approach due to their fast acquisition speed and compact size. 
However, a demosaicking algorithm is required to fully recover the spatial and spectral information of the snapshot images.
Most state-of-the-art demosaicking algorithms require ground-truth training data with paired snapshot and high-resolution hyperspectral images, but such imagery pairs with the exact same scene are physically impossible to acquire in intraoperative settings.
In this work, we present a fully unsupervised hyperspectral image demosaicking algorithm which only requires exemplar snapshot images for training purposes.\hfill~ 

\textbf{Methods:} We regard hyperspectral demosaicking as an ill-posed linear inverse problem which we solve using a deep neural network.
We take advantage of the spectral correlation occurring in natural scenes to design a novel inter spectral band regularisation term based on spatial gradient consistency.
By combining our proposed term with standard regularisation techniques and exploiting a standard data fidelity term, we obtain an unsupervised loss function for training deep neural networks, which allows us to achieve real-time hyperspectral image demosaicking.\hfill~ 

\textbf{Results:} Quantitative results on hyperspetral image datasets show that our unsupervised demosaicking approach can achieve similar performance to its supervised counter-part, and significantly outperform linear demosaicking.
A qualitative user study on real snapshot hyperspectral surgical images confirms the results from the quantitative analysis.\hfill~

\textbf{Conclusion:} Our results suggest that the proposed unsupervised algorithm can achieve promising hyperspectral demosaicking in real-time thus advancing the suitability of the modality for intraoperative use.\hfill~}




\maketitle

\section{Introduction} \label{sec:intro}
Hyperspectral Imaging (HSI) is a technique that captures and processes spectral data distributed across a large number of wavelengths. It provides a non-contact, non-ionising and non-invasive solution suitable for many medical applications \citep{lu2014medical,shapey2019intraoperative,clancy2020surgical}. HSI can provide information beyond what human vision can observe, such as tissue perfusion, oxygen saturation, and other diagnostic measurements \citep{holmer2018hyperspectral}. 
Hence, it facilitates important medical tasks such as tissue differentiation and characterisation. Depending on the number of bands, hyperspectral imaging may also be called multispectral imaging, but in this work we will refer to hyperspectral imaging for consistency.

\begin{figure}[tb]
\centering
    \includegraphics[height=0.28\textwidth]{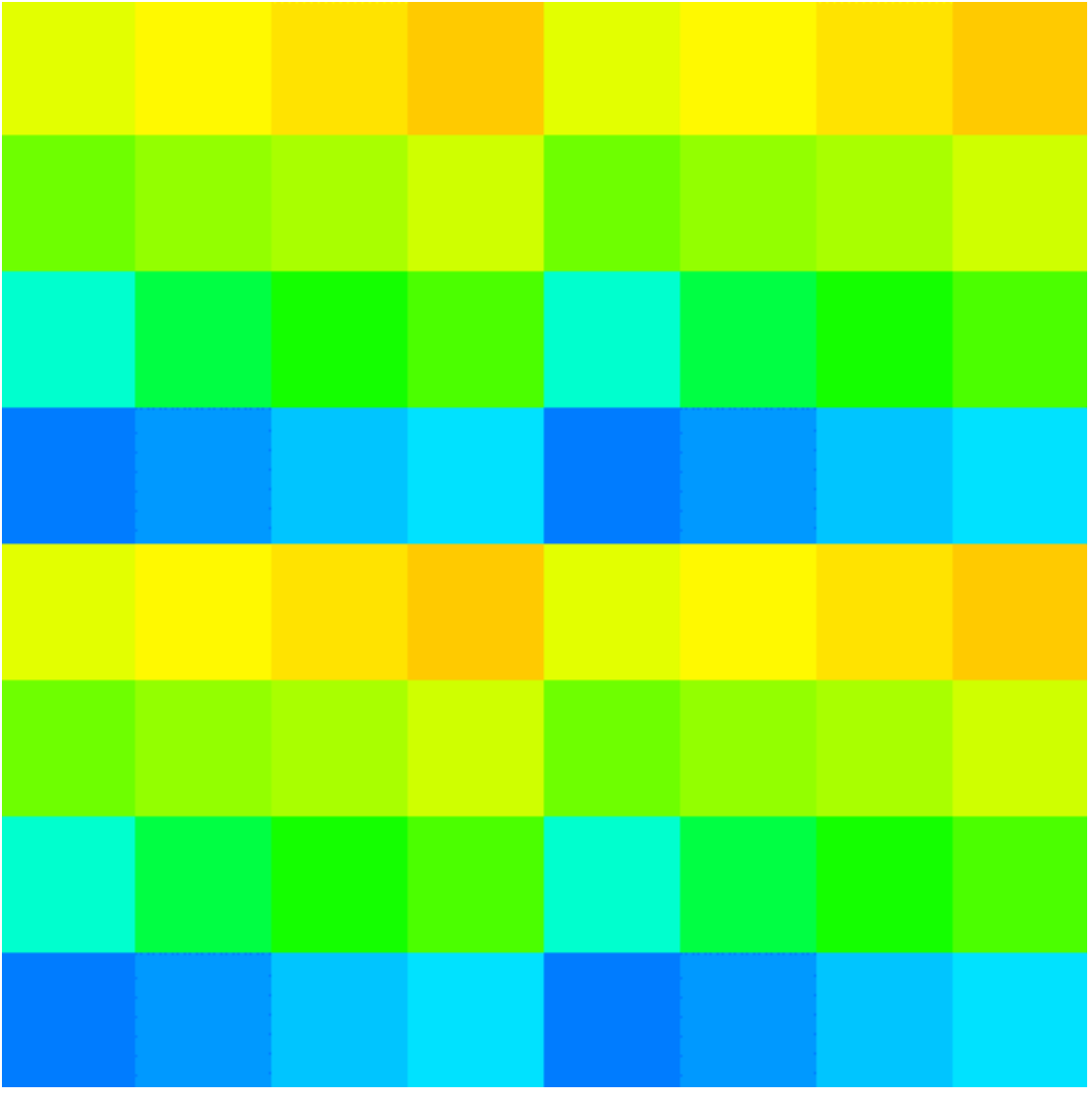}
    \includegraphics[height=0.28\textwidth]{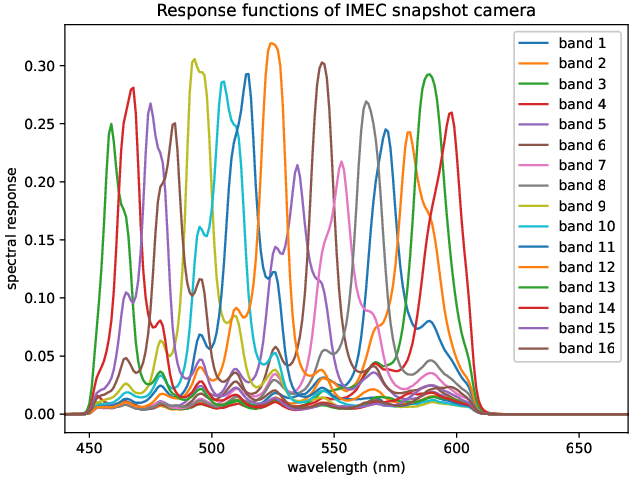}
    \includegraphics[height=0.28\textwidth]{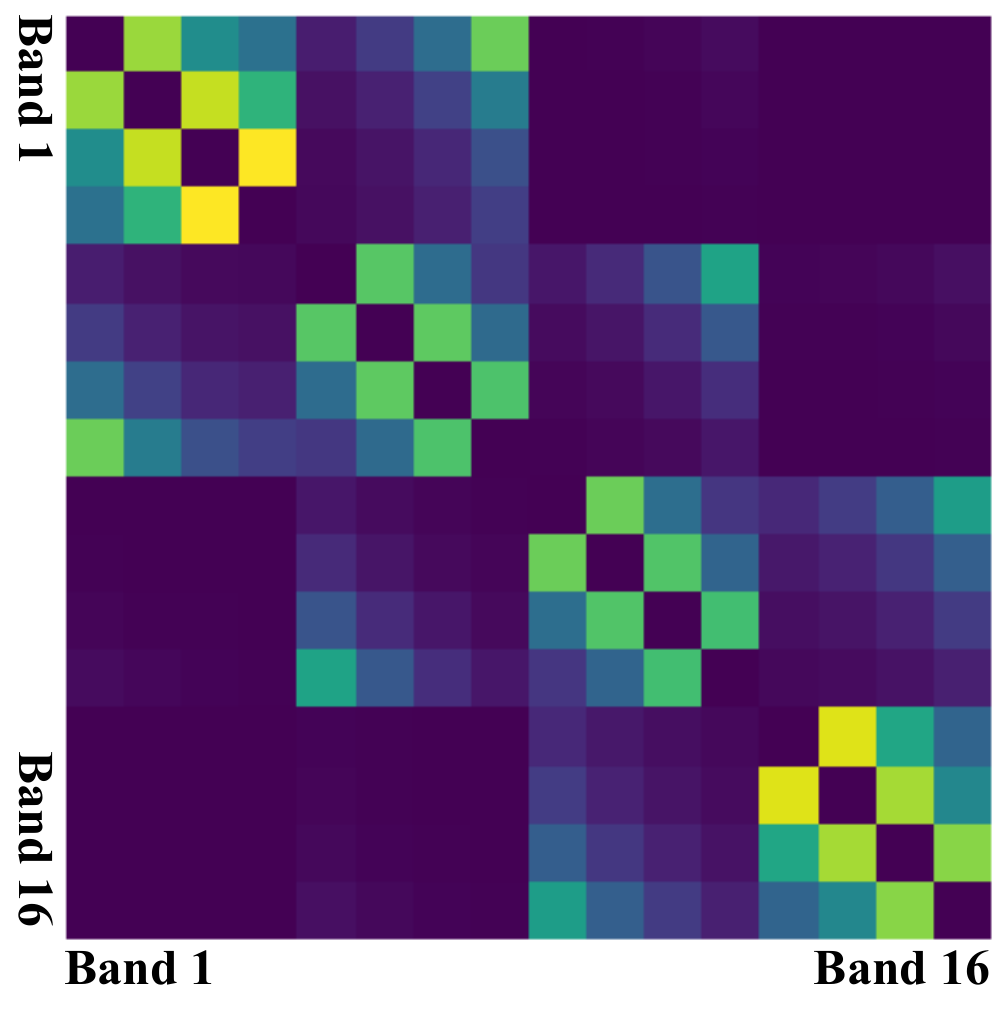}
\caption{\label{fig:sensor}
(Left) $4 \times 4$ MSFA for an IMEC snapshot camera, the colour of each pixel correlates to the perceived colour of a human observer. 
(Middle) Spectral responses of all 16 sensors on the MSFA of an IMEC snapshot camera.
(Right) Wasserstein metric heatmap measuring distances between different spectral responses of the sensors on an IMEC snapshot camera.
}
\end{figure}

Snapshot hyperspectral imaging is a promising technique which can capture hyperspectral images in real-time. Snapshot mosaic cameras are a common type of snapshot hyperspectral camera which employ multi-spectral filter array (MSFA) to acquire multi-spectral data in a single exposure.
In MSFA cameras the $n \times n$ sensor arrays are arranged in a repeating pattern similar to the $2 \times 2$ Bayer filter arrays on RGB cameras 
(\figref{fig:sensor}, left)
and are thus capable of obtaining a maximum of $n^2$ bands instantly. However, it achieves real-time multi-spectral data acquisition at the cost of reducing both spatial and spectral resolution.
Efficient hyperspectral demosaicking algorithms are thus required to fully restore the spatial and spectral resolution from the snapshot images.
More details on hyperspectral imaging techniques and snapshot mosaic imaging can be found in \citep{li2022deep}.

Traditionally, demosaicking algorithms were developed using interpolation-based methods or statistics-based techniques \citep{yu2006colour,eismann2004application}, but these methods may still suffer from colour artifacts and blurriness.
Recent deep-learning based algorithms have been developed for efficient and accurate image super-resolution and demosaicking tasks. Deep neural networks such as SRCNN \citep{dong2014learning}, EDSR \citep{lim2017edsr} and RNAN \citep{zhang2019residual} have demonstrated their performances on RGB image super-resolution tasks, and thus similar methods have been extended to process hyperspectral images \citep{mei2017hyperspectral, dijkstra2019hyperspectral}. \citep{arad2022ntire} introduced several state-of-the-art learning-based hyperspectral demosaicking algorithms of natural scenes in NTIRE 2022 Spectral Demosaicking Challenge. The leading contestants include Enhanced HAN \citep{niu2020single}, NLRAN \citep{arad2022ntire} and Res2-Unet based methods~\citep{song2022hyperspectral}. Our previous work \citep{li2022deep} also demonstrated the use of a synthetic surgical HSI dataset and deep-learning models for developing hyperspectral demosaicking algorithms suitable for intraoperative surgical guidance tasks.

However, most deep-learning based demosaicking algorithms rely on a large number of high-resolution HSI data as the ground truth for model training. Publicly available medical hyperspectral datasets such as HELICoiD \citep{fabelo2019helicoid} and ODSI \citep{hyttinen2020oral} involve large line-scan or spectral-scan HSI systems to obtain high-resolution hyperspectral data, and the acquisition speed is slow.
Consequently, these imaging systems are not ideal for intraoperative use. Fortunately, \citep{ebner2021intraoperative} demonstrated that the acquisition of intraoperative snapshot mosaic images is less challenging as its compact imaging system can be seamlessly integrated into a standard surgical workflow. 

This paper presents an unsupervised-learning-based HSI demosaicking algorithm which uses only snapshot mosaic images and does not require corresponding high-resolution images for training. 
A demosaicking loss function is proposed based on a novel spatial gradient consistency regularisation technique combined with traditional regularisation methods including Tikhonov regularisation and total variation. 
The proposed algorithm has been tested with 3 different deep neural networks on 3 different datasets. Quantitative measures have been performed to compare the unsupervised algorithm against linear demosaicking and supervised training, and a qualitative user study was conducted to validate the proposed algorithm on a medical HSI dataset.

\section{Materials and methods}\label{sec:method}

\subsection{Demosaicking as an ill-posed linear inverse problem}

\myparagraph{Problem formulation}
Hyperspectral image demosaicking involves recovering the fully sampled hyperspectral image $I \in \mathbb{R}^{X \times Y \times C}$ from a snapshot image $I^s \in \mathbb{R}^{X \times Y}$,
where $X$ and $Y$ are the spatial dimensions and $C$ is the number of spectral bands.
The relationship between $I$ and $I^s$ can be expressed through a linear degradation operator $\mathcal{D}$:
\begin{equation} \label{eq:snapshot}
    I^s=\mathcal{D}(I)
\end{equation}

For a typical MSFA arrangement as shown in \figref{fig:sensor} (left), $\mathcal{D}$ can be simply expressed as a selection matrix containing only 0 and 1, thereby mapping the pixel values of $I^s$ from $I$.
In other words, for each spatial location $(x,y)$, there is a single corresponding spectral band $c_{x,y}$ such that
$I^s(x,y) = I(x,y,c_{x,y})$
The inverse problem corresponding to \eqref{eq:snapshot} is ill-posed because of the highly ill-conditioned selection operator $\mathcal{D}$.
Therefore, appropriate regularisation is required.
A classical inverse problem approach would aim at solving for
\begin{equation} \label{eq:inv}
    \hat{I} = \arg\min_I \big[ \mathcal{L}(I^s,\mathcal{D}(I)) + \lambda \mathcal{R}(I) \big]
\end{equation}
where $\mathcal{L}(I^s,\mathcal{D}(I))$ is the data fidelity term that measures the differences between the the known snapshot image $I^s$ and the subsampling of the unknown fully-sampled hyperspectral image $I$.
$\mathcal{R}$ represents the regularisation terms. $\lambda$ is the regularisation factor that determines the trade-off between the data fidelity and regularisation.

Translating this into an unsupervised machine learning setting, we now seek to optimise for the parameters $\theta$ of a deep neural network $f_{\theta}$ mapping a snapshot mosaic input $I^s$ to a fully-sampled hyperspectral image $f_{\theta}(I^s)$:
\begin{equation} \label{eq:inv-ml}
    \hat{\theta} = \arg\min_\theta \mathbb{E}_{I^s} \big[ \mathcal{L}(I^s,\mathcal{D}(f_{\theta}(I^s))) + \lambda \mathcal{R}(f_{\theta}(I^s)) \big]
\end{equation}
where the expectation $\mathbb{E}_{I^s}$ is to be considered as being taken over an empirical distribution defined by a training set of snapshot mosaic images (with no need for ground truth).

\myparagraph{Spatial gradient consistency regularisation}
Regularisation terms in \eqref{eq:inv-ml} aim at incorporating prior information about the problem being solved.
In our case, all spectral bands are imaging the same physical scene. We also observe that the spectrum of natural objects and biological tissues present with specific characteristics such as continuity and smoothness.
Additionally, the response functions corresponding to the different spectral bands
as shown in \figref{fig:sensor} (middle) shares significant spectral overlap.
It is thus expected that our spectral bands will exhibit substantial correlation.
Inter-spectral band correlation was notably demonstrated empirically for RGB images in \citep{gunturk2002color}.
However, while correlation is expected, assuming a simple linear relationship would make for too crude an approximation.

Here, inspired by image similarity metrics that exploit image gradients for multimodal image registration where non-trivial correlation across the imaging modalities is expected \citep{haber2006intensity}, we propose to promote correlation between the spatial gradients of the individual spectral bands in our reconstructions.
Let $c_1$ and $c_2$ be the indices of two spectral bands of interest, with $I^{c}=I(\cdot,\cdot,c)$, and $c\in(c_1,c_2)$ the corresponding spectral band images.
For simplicity, we make use of forward differences to compute spatial gradients:
$\nabla_x I^c(x,y)=I^c(x+1,y)-I^c(x,y)$ and $\nabla_y I^c(x,y)=I^c(x,y+1)-I^c(x,y)$.
We propose to consider the correlation coefficient between the spatial gradients as a regularisation:
\begin{equation} \label{eq:corr_pair}
    \mathcal{R}^{c_1,c_2}_{\corr}(I) =
    - \corr(\nabla_x I^{c_1}, \nabla_x I^{c_2})
    - \corr(\nabla_y I^{c_1}, \nabla_y I^{c_2})
\end{equation}

Given $C$ spectral bands, $C^2$ pairwise comparisons are possible.
However, the strength of the correlation is not expected to be the same for all pairs of bands.
Indeed, two bands with close spectral peaks should lead to higher correlation than two bands with further peaks.
Given the complex structure of the spectral response functions shown in \figref{fig:sensor} (middle), we propose to weight the contribution of each pair of spectral band according to the Wasserstein distance $W_{c_1,c_2}$ between the spectral response functions of the two bands:
\begin{equation} \label{eq:corr}
    \mathcal{R}_{\corr}(I) =
    \sum_{c_1 \neq c_2}
    e^{-\frac{W_{c_1,c_2}}{\tau}} ~
    \mathcal{R}^{c_1,c_2}_{\corr}(I)
\end{equation}
where the negative exponential mapping with temperature scaling $\tau$ allows to control the relative importance of each pair.
The exponential Wasserstein distance gives an indication of how closely the spectral responses of the two bands might be correlated, as shown in the heatmap in \figref{fig:sensor} (right), where lighter colour means the two spectral bands are closer. 
By strengthening the correlation between the spatial gradient maps of different spectral bands we expect to enhance the sharp edges and contours.

\myparagraph{Other regularisation terms}
Tikhonov regularisation is a common method for ill-conditioned problems. It can be characterised as:
\begin{equation} \label{eq:tik}
    \mathcal{R}_{\textrm{Tik}}(I)=\|\bm{\Gamma} \cdot I\|_2^2
\end{equation}
Here, we choose to use the
Laplacian matrix as the Tikhonov matrix $\bm{\Gamma}$ to deal with potential high-frequency artifacts introduced during the super-resolution process.
While Tikhonov regularisation can effectively eliminate undesirable outliers and led to smooth images, it also has the potential risk of applying too much smoothness and erasing all sharp edges and contours, which is harmful for recovering details in the images. 

Total variation is another term which is able to preserve edges while regularising solutions of the inverse problem:
\begin{equation} \label{eq:tv}
    \mathcal{R}_{\textrm{TV}}(I) =
    \|\nabla_x I\|_1 + \|\nabla_y I\|_1
\end{equation}

By combining our proposed spatial gradient consistency term with
Tikhonov and total variation regularisation,
we obtain the regularisation term $\mathcal{R}$ in \eqref{eq:inv} using $\lambda_{\textrm{Tik}}$, $\lambda_{\textrm{TV}}$ and $\lambda_{\corr}$ as weighting factors for individual terms:
\begin{equation} \label{eq:inv2}
    \mathcal{R}(I) =
    \lambda_{\textrm{Tik}} \mathcal{R}_{\textrm{Tik}}(I)
    + \lambda_{\textrm{TV}} \mathcal{R}_{\textrm{TV}}(I)
    + \lambda_{\corr} \mathcal{R}_{\corr}(I)
\end{equation}

\subsection{Image demosaicking pipeline}

\begin{figure}
\centering
\includegraphics[width=0.9\textwidth]{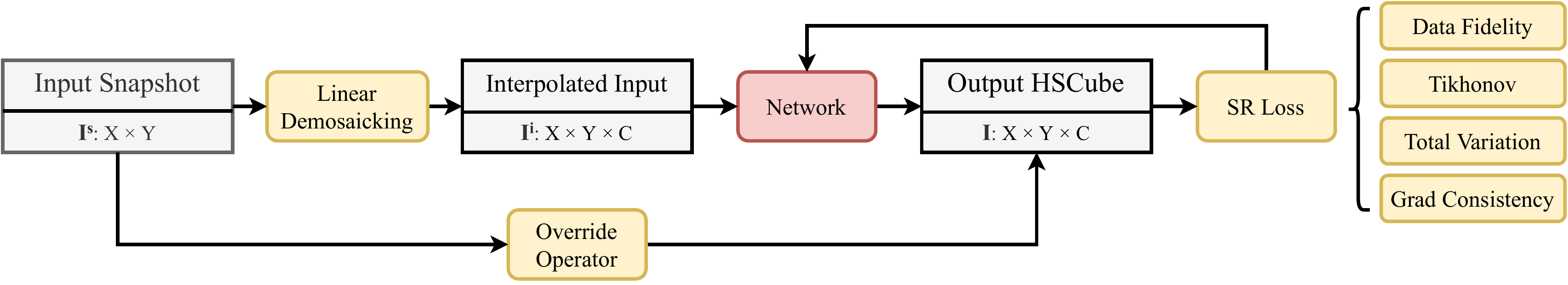}
\caption{The pipeline of the proposed unsupervised demosaicking algorithm.} \label{fig:pipeline}
\end{figure}

\figref{fig:pipeline} depicts the general pipeline of our proposed algorithm using deep neural networks for hyperspectral image demosaicking problems. It starts from the input snapshot mosaic images where bilinear interpolation based demosaicking can be applied to recover the spatial and spectral dimension of the images.
The linearly interpolated images serve as the input of the network to generate refined demosaicking results. Most deep neural networks for image super-resolution or demosaicking can be integrated into this pipeline, such as U-Net \citep{ronneberger2015unet}, EDSR \citep{lim2017edsr} and Res2-Unet \citep{song2022hyperspectral}.

Aside from the network,
given that the measured pixels in the original snapshot $I^s$ should be equal to the corresponding pixels in the demosaicked hypercube $I$,
we propose to include an overriding operator which applies the pixel values from $I^s$ to their corresponding position in $I$.
This forces the data fidelity term $\mathcal{L}$ in \eqref{eq:inv} to be always 0 irrespective of the metric we choose.
Based on the output images from the network with the overridden snapshot pixels, the Tikhonov regularisation, total variation and the spatial gradient consistency regularisation terms are calculated and minimised using gradient descent, and the parameters in the networks are updated.

\subsection{Source datasets} \label{subsec:datasets}
To experiment the proposed demosaicking algorithm, three hyperspectral imaging datasets are used in this work, which will be presented in this section.

\myparagraph{HELICoiD}
\citep{fabelo2019helicoid} presented a publicly available in-vivo hyperspectral human brain image dataset within the European project HELICoiD (HypErspectraL Imaging Cancer Detection). The hyperspectral images in this dataset were acquired using a line-scan hyperspectral camera system capable of capturing high spectral-resolution hypercubes during neurosurgical operations. The dataset contains 36 images in the Visual and Near Infrared (VNIR) range from 400nm to 1000nm. We applied the same method described in \citep{li2022deep} to perform white balancing, and then simulated snapshot mosaic images and their corresponding high-resolution demosaicked hypercubes using spectral response functions of a real hyperspectral snapshot camera.

\myparagraph{ARAD\_1K}
With the NTIRE 2022 Spectral Demosaicking Challenge, \citep{arad2022ntire} provided
1000 hyperspectral images of natural scenes with 16 spectral bands ranging from 400nm to 1000nm.
The snapshot images were simulated following a $4 \times 4$ MSFA pattern. There were 950 hyperspectral images for training, where the simulated snapshot images and their corresponding ground truth images were both provided. The other 50 images were for testing, but the ground truth was not publicly available, so we separated 50 images out from the 950 training set for testing.

\myparagraph{NeuroHSI}
NeuroHSI is an actively running, NIHR funded, single centre prospective observational study assessing the intra-operative capabilities of a $4 \times 4$, 16 band visible range snapshot mosaic camera (IMEC CMV2K-SSM4X4-VIS) to differentiate between pathological tissue and healthy brain tissue, as well as to evaluate custom made algorithms capable of correlating information from specific bands to tissue oxygenation measurements. Phase one of this study has now been completed and video hyperspectral data from two brain metastases, two gliomas (WHO grades 2-4), one meningiomas, one vestibular schwannoma, one cerebral aneurysm and one cerebral arteriovenous malformation has been collected. 
150 snapshot images with minor motion blur or out-of-focus blur were manually selected from the video data of the 8 patients, where 90 images from 4 patients are reserved for training, 30 images from 2 patients reserved for validation and 30 from the remaining 2 patients for testing.

\subsection{Implementation details}
Our proposed algorithm was implemented with PyTorch and tested on all three datasets described in Section \ref{subsec:datasets}. For the HELICoiD dataset, synthetic snapshot images and their corresponding high-resolution hypercubes were simulated using sensor information from the snapshot camera IMEC CMV2K-SSM4X4-VIS. The dataset was divided into 3 groups: 24 images acquired from 15 different patients as the training set, 6 images from 4 patients as the validation set, and the remaining 6 images from 3 patients as the test set. For the ARAD\_1K dataset, the original raw snapshot data were simulated with an unknown exposure setting.
Recovering such an unknown exposure is not the primary focus for our experiment. Therefore, new snapshot images were simulated using the ground truth hypercubes and the MSFA simulation algorithm provided by the organiser. The dataset was also divided into 3 groups: 720 images for training, 180 for validation and 50 for testing. 

As both the HELICoiD and ARAD\_1K datasets have high-resolution hypercubes as ground truths, the U-Net, EDSR and Res2-Unet models were trained in both a supervised and an unsupervised manner. For supervised training, the models were all trained using the Mean Relative Absolute Error (MRAE) Loss as described in \citep{song2022hyperspectral}. For unsupervised training, the regularisation terms described in \eqref{eq:inv2} were used as the loss function, and the models were trained with only the simulated snapshot images as inputs. The regularisation factors in \eqref{eq:inv2} were set to $\lambda_{\textrm{Tik}}=1$, $\lambda_{\textrm{TV}}=10^{-3}$ and $\lambda_{\corr}=1$ respectively, and the temperature scaling $\tau$ in \eqref{eq:corr} was set to 0.1. 
Details on the parameter selection and the ablation study can be found in the supplementary material.
Random flipping and rotation were not performed because they can disrupt the MSFA pattern of the snapshot images. Therefore, random divisible spatial cropping were performed where the position and size of the crop were all divisible by the size of the mosaic.
The network models were trained using the Adam optimiser with $\beta_1=0.5$ and $\beta_2=0.99$ and a batch size of 4. The initial learning rate was set to $1 \times 10^{-4}$. Results were quantitatively evaluated based on 3 metrics, including Structural Similarity (SSIM), Peak Signal-to-Noise Ratio (PSNR) and Spectral Angle Mapper (SAM) \citep{kruse1993spectral}.

The 150 image frames selected from the NeuroHSI video dataset were all acquired from an IMEC CMV2K-SSM4X4-VIS camera, and there are no ground truth high-resolution hypercubes, so the experiment only involves unsupervised training. 90 snapshot image frames from 4 patients were used for training, and 30 images from 2 patients for both validation and testing. Res2-Unet was adopted for the proposed algorithm, and the parameters used for training on NeuroHSI dataset remains the same as the HELICoiD and ARAD\_1K dataset. The results were evaluated qualitatively by a user study which will be described in Section \ref{subsec:user-study}.

\section{Results}\label{sec:result}

\begin{table}[bt!]
\begin{center}
\caption{Comparison of demosaicking accuracy between linear demosaicking and different networks with supervised and unsupervised training setup on HELICoiD and ARAD\_1K datasets.}
\label{tab:helicoid-results}
\resizebox{\textwidth}{!}{
\begin{tabular}{ccccccc}
\toprule
Dataset     & Method                        & Network   & SSIM              & PSNR            & SAM \\
\midrule
\multirow{7}{*}{HELICoiD} & Linear          & -         & $0.969 \pm 0.011$ & $32.5 \pm 2.48$ & $0.123 \pm 0.051$ \\
\cmidrule{2-6}
            & \multirow{3}{*}{Supervised}   & UNet      & $0.991 \pm 0.005$ & $38.7 \pm 3.06$ & $0.057 \pm 0.025$ \\
            &                               & EDSR      & $0.994 \pm 0.004$ & $42.3 \pm 3.70$ & $0.035 \pm 0.013$ \\
            &                               & Res2-Unet & $\mathbf{0.995 \pm 0.004}$ & $\mathbf{42.9 \pm 3.99}$ & $\mathbf{0.032 \pm 0.012}$ \\
\cmidrule{2-6}
            & \multirow{3}{*}{Unsupervised} & UNet      & $0.991 \pm 0.004$ & $38.6 \pm 3.43$ & $0.057 \pm 0.024$ \\
            &                               & EDSR      & $\mathbf{0.993 \pm 0.004}$ & $\mathbf{39.9 \pm 3.70}$ & $\mathbf{0.046 \pm 0.019}$ \\
            &                               & Res2-Unet & $0.992 \pm 0.005$ & $39.7 \pm 3.66$ & $0.051 \pm 0.019$ \\
\midrule
\multirow{7}{*}{ARAD\_1K} & Linear          & -         & $0.956 \pm 0.039$ & $33.4 \pm 4.16$ & $0.055 \pm 0.022$ \\
\cmidrule{2-6}
            & \multirow{3}{*}{Supervised}   & UNet      & $0.978 \pm 0.022$ & $38.4 \pm 3.91$ & $0.041 \pm 0.016$ \\
            &                               & EDSR      & $0.993 \pm 0.007$ & $43.1 \pm 3.74$ & $0.025 \pm 0.009$ \\
            &                               & Res2-Unet & $\mathbf{0.998 \pm 0.002}$ & $\mathbf{48.7 \pm 3.88}$ & $\mathbf{0.015 \pm 0.006}$ \\
\cmidrule{2-6}
            & \multirow{3}{*}{Unsupervised} & UNet      & $0.977 \pm 0.020$ & $38.1 \pm 3.62$ & $0.041 \pm 0.015$ \\
            &                               & EDSR      & $\mathbf{0.989 \pm 0.013}$ & $41.0 \pm 4.07$ & $0.031 \pm 0.013$ \\
            &                               & Res2-Unet & $0.988 \pm 0.017$ & $\mathbf{41.3 \pm 4.42}$ & $\mathbf{0.031 \pm 0.012}$ \\
\botrule
\end{tabular}
}
\end{center}
\end{table}

\subsection{Quantitative evaluation} \label{subsec:quantitative}

The quantitative results of the demosaicked hypercubes on both HELICoiD and ARAD\_1K datasets are shown in \autoref{tab:helicoid-results}. Paired T-test was performed to compare against the performance of two demosaicking methods. For both datasets, the supervised training of Res2-Unet achieved the highest demosaicking accuracy. The supervised EDSR results did not show statistical differences compared to Res2-Unet at a significant level of 0.05 on the HELICoiD dataset, with p-values of 0.35, 0.34 and 0.30 for SSIM, PSNR and SAM respectively. However, on the ARAD\_1K dataset the p-values of $<10^{-5}$ for all 3 metrics indicates that Res2-Unet outperforms EDSR significantly. 

The demosaicking results of the proposed unsupervised method on Res2-Unet are significantly lower than the supervised method with p-values of 0.040, 0.016, 0.007 on the 3 metrics on HELICoiD dataset, and p-values of close to 0 on ARAD\_1K dataset, showing that our proposed method cannot match state-of-the-art supervised demosaicking methods when ground truths are provided. However, when comparing supervised and unsupervised EDSR results, the p-values of 0.17, 0.06 and 0.07 on the HELICoiD dataset indicates that our proposed method can still reach similar performance as a supervised method. On the ARAD\_1K dataset, although the unsupervised EDSR performs significantly lower than supervised EDSR with p-values of 0.02, 0.0001 and 0.0005, it still outperforms the supervised U-Net significantly with p-values of $<10^{-5}$ for all 3 metrics. In both datasets, all supervised and unsupervised results significantly outperform linear demosaicking with p-values close to 0.

The speed of our proposed demosaicking algorithm depends on the choice of network.
For a single image of size $512 \times 480$ from the ARAD\_1K dataset, the inference times for UNet, EDSR and Res2-Unet are around 0.009s, 0.006s and 0.010s respectively with NVIDIA RTX 3080 Ti. This demonstrates that when combining a suitable neural network and computing hardware, our proposed algorithm can achieve high quality hyperspectral demosaicking in real-time.

\begin{figure}
\includegraphics[width=\textwidth]{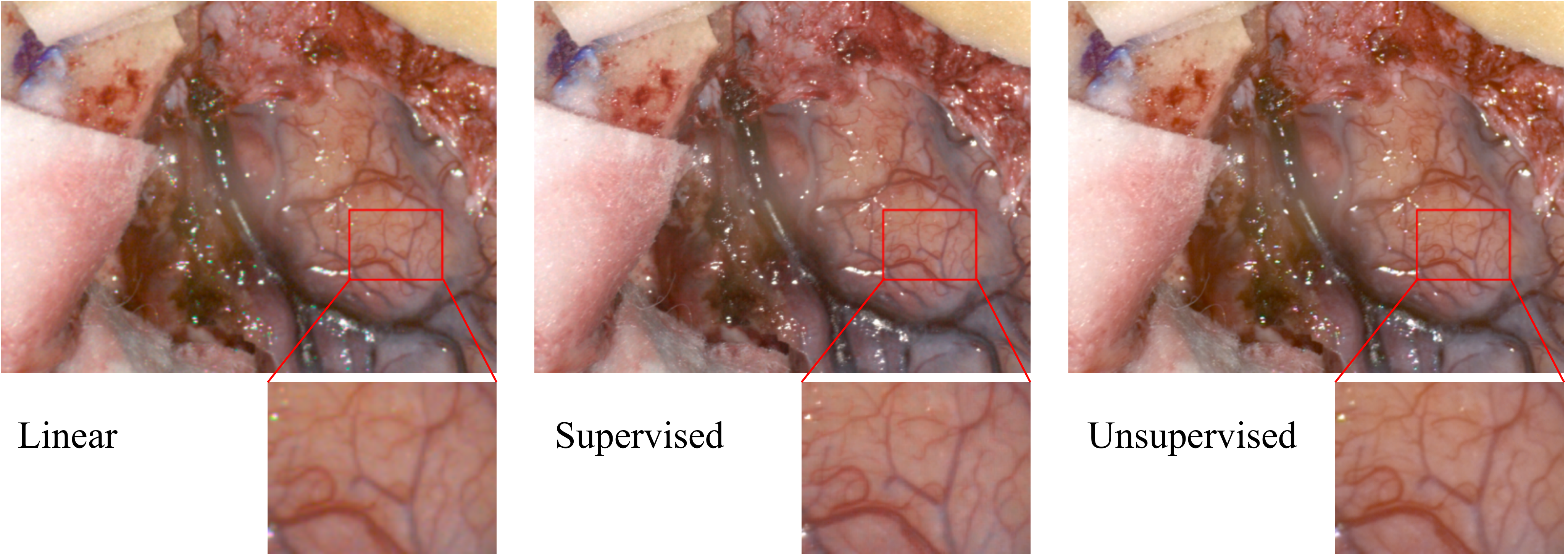}
\caption{Comparison between different demosaicking methods on an example NeuroHSI test image. The reconstructed sRGB images are converted from the demosaicked hyperspectral data following the method described in \citep{li2022deep}.} \label{fig:neurohsi-results}
\end{figure}

\subsection{Qualitative evaluation and user study} \label{subsec:user-study}

As there is no ground truth data for the NeuroHSI dataset, a qualitative user study was conducted to evaluate the demosaicked results of the NeuroHSI dataset.
The user study was conducted using forced-choice pairwise comparison \citep{mantiuk2012image}.
\figref{fig:neurohsi-results} illustrates the pseudo-sRGB reconstructions of an example NeuroHSI patient image tested using three methods: linear demosaicking (L), supervised Res2-Unet model trained from HELICoiD dataset (SL) and the unsupervised Res2-Unet model trained from NeuroHSI training set (UL).
30 test images were included in the user study, each tested with the three methods (L, SL, UL).
There are thus 90 questions in total, each containing two images of the same scene with 2 different demosaicking methods. These questions were divided into 3 separate surveys, each containing 30 questions. Participants were randomly assigned to answer one of 3 surveys and asked to choose the image with better quality for each question (pair of images) without any knowledge of which demosaicking method was used.
The participants of this survey were all neurosurgical experts with 2 to 15 years of experience. We received 12 responses in total that are summarised in \autoref{tab:user-study}. We applied the Bradley-Terry model \citep{bradley1952rank} to rank the demosaicking methods, which gives the estimated preference scale of $\pi=(0.050, 0.445, 0.505)$ for L, SL and UL respectively. This indicates that the experts considered the images recovered from our proposed demosaicking method to have similar quality as the images from a supervised model, with the baseline linear demosaicking the least favourable method. More details can be found in the supplementary material.

\begin{table}[bt]
\centering
\caption{Number of votes received for each demosaicking method in all pairwise comparisons in the image quality assessment survey.}
\label{tab:user-study}
\begin{tabular}{cc|cc|cc}
\toprule
Linear & Supervised & Linear & Unsupervised & Supervised & Unsupervised \\
\midrule
13 & 107 & 10 & 110 & 57 & 63 \\
\botrule
\end{tabular}
\end{table}

\section{Conclusion} \label{sec:conclusion}
In this work, we have presented a novel unsupervised approach for medical hyperspectral image demosaicking. The proposed algorithm does not rely on high-resolution medical hyperspectral data which are hard to acquire in a surgical environment, but instead only snapshot mosaic images are required, which are much easier to capture. The combination of Tikhonov regularisation, total variation and spectral correlation regularisation has been adopted for unsupervised network training, and the results were tested both quantitatively and qualitatively, showing convincing results over basic linear demosaicking, and comparable results against supervised demosaicking methods, thus proving its capability for real-time intraoperative surgical application.

\backmatter

\bmhead{Supplementary information}
For more information regarding details of the ablation study, quantitative metrics, qualitative results and the user study, please refer to the supplementary document alongside this article.





\section*{Declarations}
\begin{itemize}
    \item Funding: 
    This study/project is funded by the NIHR [NIHR202114]. The views expressed are those of the author(s) and not necessarily those of the NIHR or the Department of Health and Social Care.
    This work was supported by core funding from the Wellcome/EPSRC [WT203148/Z/16/Z; NS/A000049/1].
    This project has received funding from the European Union's Horizon 2020 research and innovation programme under grant agreement No 101016985 (FAROS project).
    TV is supported by a Medtronic / RAEng Research Chair [RCSRF1819\textbackslash7\textbackslash34].
    PL is funded by China Scholarship Council.
    CH is supported by an InnovateUK Secondment Scholars Grant (Project Number 75124).
    For the purpose of open access, the authors have applied a CC BY public copyright license to any Author Accepted Manuscript version arising from this submission.
    \item Conflict of interest:
    TV and JS are co-founders and shareholders of Hypervision Surgical.
    \item Ethics approval:
    All procedures within this study involving human subjects were in accordance with both the institutional and regional ethical committee (REC reference 22/LO/0046, IRAS 284230) and with the 1964 Helsinki declaration and its later amendments. 
    \item Informed consent:
    Informed consent was obtained from all individual participants involved in the study.
    \item Consent for publication:
    The authors affirm that human research participants provided informed consent for publication of the images in \figref{fig:neurohsi-results}.
    
\end{itemize}

\bibliography{references}



\section*{Supplementary material}

# Supplementary material
## Spatial gradient consistency for unsupervised learning of hyperspectral demosaicking: application to surgical imaging

Peichao Li[1*], Muhammad Asad[1], Conor Horgan[1], Oscar MacCormac[1,2], Jonathan Shapey[1,2] and Tom Vercauteren[1]

[1]School of Biomedical Engineering & Imaging Sciences, King's College London, London, UK.
[2] Department of Neurosurgery, King's College Hospital NHS Foundation Trust, London, UK.

*Corresponding author(s). E-mail(s): peichao.2.li@kcl.ac.uk;

# 1 Evaluation Metrics

The evaluation metrics used in our quantitative analysis of the demosaicking results include Structural Similarity index (SSIM), Peak Signal-to-Noise Ratio (PSNR) and Spectral Angle Mapper (SAM) [1]. Given the ground truth hyperspectral image $I \in \mathbb{R}^{X \times Y \times C}$ and the demosaicked hyperspectral image $\hat{I} \in \mathbb{R}^{X \times Y \times C}$, the SSIM can be calculated based on the luminance term $l(I, \hat{I})$, the contrast term $c(I, \hat{I})$ and the structural term $s(I, \hat{I})$:

$$l(I, \hat{I}) = \frac{2\mu_I \mu_{\hat{I}} + C_1}{\mu_I^2 + \mu_{\hat{I}}^2 + C_1} \tag{1}$$

$$c(I, \hat{I}) = \frac{2\sigma_I \sigma_{\hat{I}} + C_2}{\sigma_I^2 + \sigma_{\hat{I}}^2 + C_2} \tag{2}$$

$$s(I, \hat{I}) = \frac{\sigma_{I\hat{I}} + C_3}{\sigma_I \sigma_{\hat{I}} + C_3} \tag{3}$$

where $\mu_I$, $\mu_{\hat{I}}$, $\sigma_I$, $\sigma_{\hat{I}}$ and $\sigma_{I\hat{I}}$ represent the means, standard deviations and the cross covariance of the hyperspectral images $I$ and $\hat{I}$. $C_1$, $C_2$ and $C_3$ are

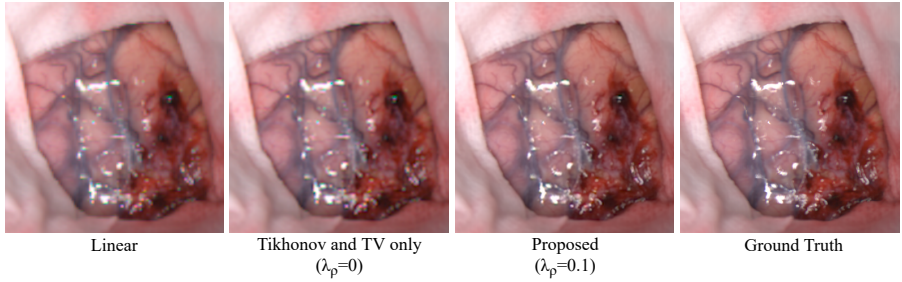

| Linear | Tikhonov and TV only ($\lambda_p$=0) | Proposed ($\lambda_p$=0.1) | Ground Truth |

**Fig. 1** The effect of the proposed gradient consistency regularisation term on the demosaicked image.

constants to ensure stability when the denominator becomes 0. SSIM is then given by:

$$SSIM(I, \hat{I}) = [l(I, \hat{I})]^\alpha \cdot [c(I, \hat{I})]^\beta \cdot [s(I, \hat{I})]^\gamma \tag{4}$$

To simplify the expression, by default $\alpha = \beta = \gamma = 1$, and $C_3 = 0.5C_2$. Hence, SSIM can be simplified as:

$$SSIM(I, \hat{I}) = \frac{(2\mu_I\mu_{\hat{I}} + C_1)(2\sigma_{I\hat{I}} + C_2)}{(\mu_I^2 + \mu_{\hat{I}}^2 + C_1)(\sigma_I^2 + \sigma_{\hat{I}}^2 + C_2)} \tag{5}$$

PSNR is defined as:

$$PSNR = 10\log_{10}(\frac{MAX_I^2}{MSE}) \tag{6}$$

where $MAX_I^2$ is the maximum pixel value in $I$, and MSE is the mean-squared error which is defined as:

$$MSE = \frac{1}{XYC} \sum_{x=1}^{X} \sum_{y=1}^{Y} \sum_{c=1}^{C} (I(x,y,c) - \hat{I}(x,y,c))^2 \tag{7}$$

Finally, SAM can be calculated by:

$$SAM = \frac{1}{XY} \sum_{x=1}^{X} \sum_{y=1}^{Y} \arccos\left(\frac{\sum_{c=1}^{C} I(x,y,c) \cdot \hat{I}(x,y,c)}{\sqrt{\sum_{c=1}^{C} I(x,y,c)^2} \sqrt{\sum_{c=1}^{C} \hat{I}(x,y,c)^2}}\right) \tag{8}$$

# 2 Ablation Study

Figure 1 shows an example test image from HELICoiD [2] dataset illustrating the effect of the proposed gradient consistency regularisation term. For easy visualisation, the hyperspectral images presented in this document have all been converted into sRGB images. This can be achieved using the method described in [3], which involves first converting the spectral data to the CIE

**Table 1** PSNR results of the demosaicked images using different weighting factors for each regularisation term.

| | | $\lambda_\rho$=0 | $\lambda_\rho$=0.01 | $\lambda_\rho$=0.1 | $\lambda_\rho$=1 | $\lambda_\rho$=10 |
|---|---|---|---|---|---|---|
| $\lambda_{TV}=0$ | $\lambda_{Tik}=0$ | - | 36.33 | 36.33 | 36.33 | 36.33 |
| $\lambda_{TV}=0$ | $\lambda_{Tik}=0.1$ | 32.54 | 39.82 | 40.32 | 38.67 | 36.78 |
| $\lambda_{TV}=0$ | $\lambda_{Tik}=1$ | 32.54 | 36.23 | 39.93 | 40.35 | 38.67 |
| $\lambda_{TV}=0$ | $\lambda_{Tik}=10$ | 32.54 | 33.66 | 36.23 | 39.92 | 40.35 |
| $\lambda_{TV}=1e-4$ | $\lambda_{Tik}=0$ | 31.64 | 40.46 | 39.12 | 36.89 | 36.38 |
| $\lambda_{TV}=1e-4$ | $\lambda_{Tik}=0.1$ | 32.15 | 39.71 | 41.47 | 38.84 | 36.83 |
| $\lambda_{TV}=1e-4$ | $\lambda_{Tik}=1$ | 32.53 | 36.21 | 39.92 | 41.37 | 38.69 |
| $\lambda_{TV}=1e-4$ | $\lambda_{Tik}=10$ | 32.54 | 33.66 | 36.23 | 39.92 | 41.35 |
| $\lambda_{TV}=1e-3$ | $\lambda_{Tik}=0$ | 31.64 | 35.16 | 38.71 | 37.80 | 36.73 |
| $\lambda_{TV}=1e-3$ | $\lambda_{Tik}=0.1$ | 30.48 | 37.41 | 41.25 | 39.50 | 37.14 |
| $\lambda_{TV}=1e-3$ | $\lambda_{Tik}=1$ | 32.29 | 35.90 | 39.81 | 41.50 | 38.84 |
| $\lambda_{TV}=1e-3$ | $\lambda_{Tik}=10$ | 32.53 | 33.65 | 36.21 | 39.92 | 41.37 |
| $\lambda_{TV}=1e-2$ | $\lambda_{Tik}=0$ | 31.60 | 33.21 | 34.80 | 37.12 | 37.67 |
| $\lambda_{TV}=1e-2$ | $\lambda_{Tik}=0.1$ | 25.15 | 29.36 | 34.39 | 39.07 | 38.05 |
| $\lambda_{TV}=1e-2$ | $\lambda_{Tik}=1$ | 29.83 | 33.03 | 37.42 | 41.26 | 39.51 |
| $\lambda_{TV}=1e-2$ | $\lambda_{Tik}=10$ | 32.29 | 33.33 | 35.90 | 39.81 | 41.50 |

XYZ colour space, and then transforming the XYZ images to linear RGB images. Finally, gamma correction is applied to obtain the sRGB images.

It can be seen from Figure 1 that when the weighting factor of the proposed regularisation term $\lambda_\rho$ is set to 0, the result from using only traditional regularisation techniques is similar to a linearly demosaicked image. The proposed regularisation term strengthens the correlation between the spatial gradient maps of different spectral bands, which results in enhanced image sharpness.

A quick way to test the performance of our proposed regularisation term without training a neural network is to solve the direct inverse problem. This can be achieved by directly finding a hyperspectral image such that the sum of all regularisation terms are minimised, as expressed in Eq. (2) in the main paper. The minimisation can be achieved by common iterative methods such as Broyden–Fletcher–Goldfarb–Shanno [4] algorithm and Adam optimisation.

The values of the weighting factors for all regularisation terms, including $\lambda_{Tik}$, $\lambda_{TV}$ and $\lambda_\rho$, were determined by solving the direct inverse problem using one of the images from the HELICoiD dataset, and then calculating the PSNR of the results. Adam optimisation with initial learning rate of 0.01, $\beta_1 = 0.5$ and $\beta_2 = 0.99$ was used during the experiment for fast minimisation, and it took 500 iterations to obtain the demosaicking results. These results are shown in Table 1, where it can be seen that the highest PSNR can be achieved when $\lambda_{Tik} = 1$, $\lambda_{TV} = 1e-3$ and $\lambda_\rho = 1$ or when $\lambda_{Tik} = 10$, $\lambda_{TV} = 1e-2$ and $\lambda_\rho = 10$, which is just multiples of the former set of weights. Therefore, we chose $\lambda_{Tik} = 1$, $\lambda_{TV} = 1e-3$ and $\lambda_\rho = 1$ for all unsupervised network training.

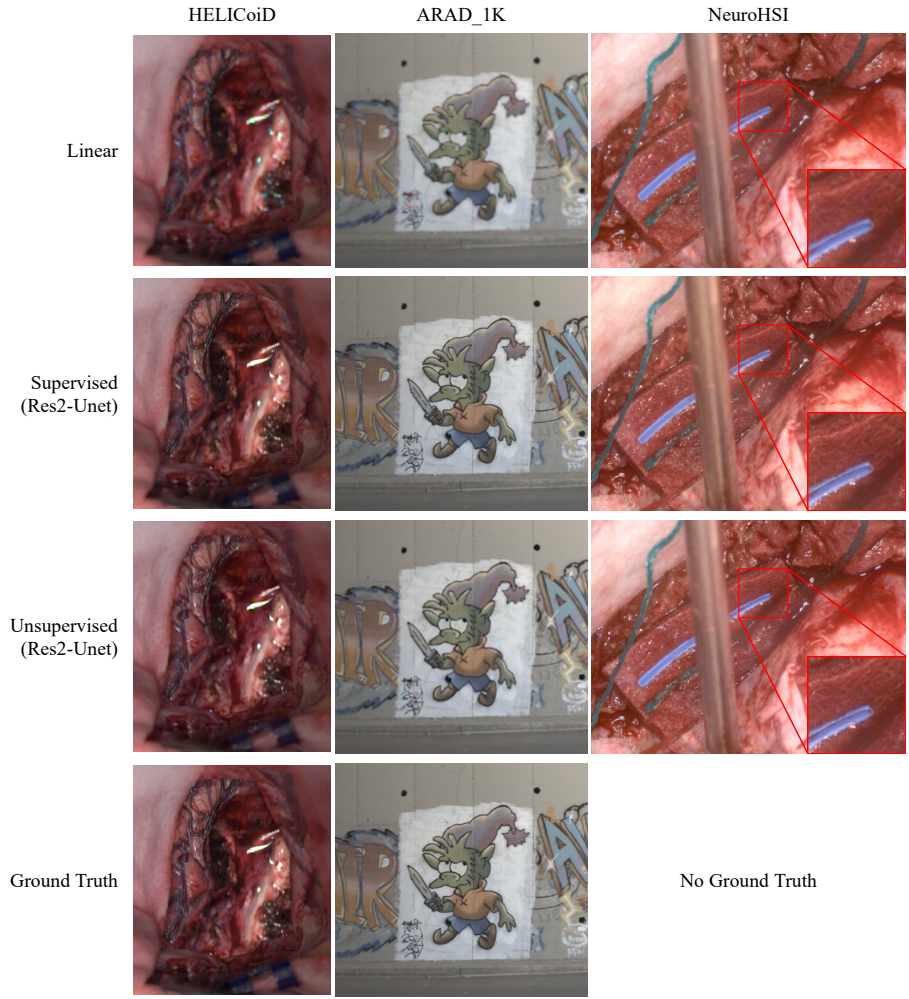

**Fig. 2** Example demosaicking results from each dataset (HELICoiD, ARAD_1K, NeuroHSI).

# 3 Additional Examples from the Results

Figure 2 shows some additional example results of linear demosaicking, supervised and unsupervised trained Res2-Unet model, as well as the ground truths respectively on HELICoiD [2], ARAD_1K [5] and NeuroHSI datasets. Since there are no high-resolution images as ground truth for the NeuroHSI dataset, training a supervised network model on this dataset is not possible. Therefore, the supervised Res2-Unet results for NeuroHSI dataset were inferred by directly using the supervised Res2-Unet model trained from HELICoiD dataset. The reason to choose the HELICoiD-trained model rather than the ARAD_1K model is that both HELICoiD and NeuroHSI are neurosurgical

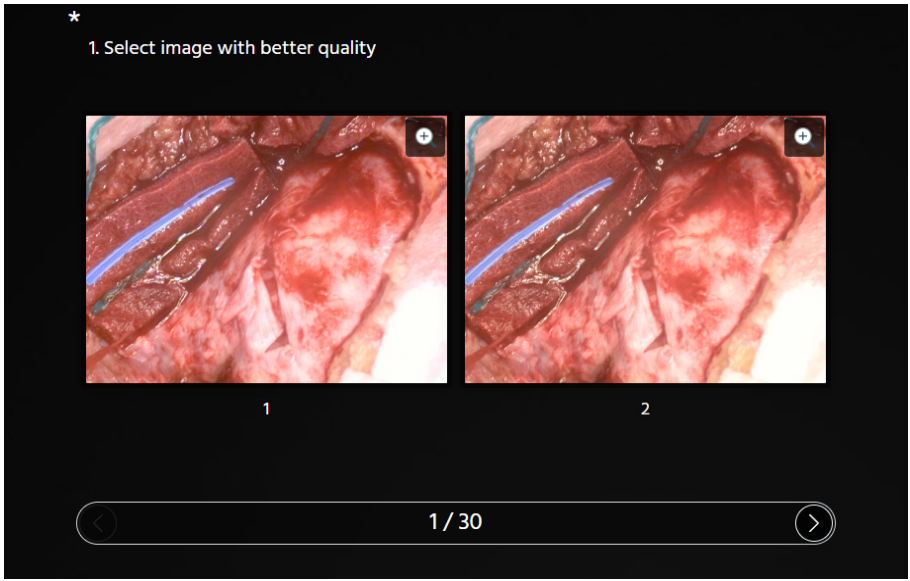

**Fig. 3** The user interface of the survey. Zoom option is provided to help the participants with observing the images in more details.

image datasets, so there is less domain gap compared to ARAD_1K which are all natural scene images. Adapting ARAD_1K-trained networks on NeuroHSI images for better results may involve methods such as transfer learning, which is not the primary focus of this work.

# 4 Additional Information on the User Study

The aim of the user study was to rank the three demosaicking methods based on the demosaicked NeuroHSI images: linear demosaicking (L), supervised training from the HELICoiD dataset (SL) and unsupervised training from the NeuroHSI dataset (UL). We chose Res2-Unet as the network to generate both supervised and unsupervised results, because from the quantitative analysis of the results on both HELICoiD and ARAD_1K datasets, supervised Res2-Unet achieved the highest demosaicking accuracy.

[6] argues that forced-choice pairwise comparison is the fastest and the most accurate type of user study for image quality assessment. Therefore, we designed a two-alternative forced-choice (2AFC) image quality survey, where observers needed to compare two images at a time and choose one with better quality without giving any rating scales. Hence, the three methods were compared by directly inferred pairwise comparison: L vs SL, L vs UL, and SL vs UL. There are 30 test images in the NeuroHSI dataset, thus the 90 demosaicked results from all three methods can form 90 pairwise comparisons as the

6     *Supplementary Material*

**Table 2** Results of the image quality assessment survey summarising the preferences for each pairwise comparison. The number refers to the number of votes that the demosaicking method in the row is preferred over the method in the column.

|      | L   | SL  | UL  |
| ---- | --- | --- | --- |
| L    | -   | 13  | 10  |
| SL   | 107 | -   | 57  |
| UL   | 110 | 63  | -   |

survey questions. It was not practical to ask each participant to make judgements on all 90 image pairs, so we divided them into 3 separate surveys, each containing 30 image pairs.

The participants of this survey were all neurosurgical experts with 2 to 15 years of experience. Each participant was randomly assigned with one of the three surveys on a website. The introductory page provided some instructions about the recommended screen size, browsers as well as the information about the survey tools. Then the two images to compare were presented to the participants, as shown in Figure 3.

We received 4 responses for each of the 3 surveys, so there are 12 responses in total. The results are summarised in Table 2, where it can be seen that when compared with linear demosaicking, the supervised demosaicking received 89.2% of the votes, and our proposed unsupervised demosaicking received 91.7% of the votes. When directly comparing the unsupervised demosaicking images against supervised demosaicking, our proposed method still received 52.5% of the votes.

Bradley-Terry model was applied to map probability of preference to scales to describe which demosaicking methods are more preferred by the experts, as suggested by [7] for image quality assessment. In the Bradley-Terry model, consider $K$ number of methods to be compared. For method $i$ and $j$, denote the probability that method $i$ wins over method $j$ as:

$$p_{ij} = \mathbb{P}(i > j) = \frac{\pi_i}{\pi_i + \pi_j} \tag{9}$$

where $\pi_i$ is the scale value indicating the preference of the method $i$. Let $w_{ij}$ be the number of votes that method $i$ is preferred over method $j$. Assume the vote from each pairwise comparison is independent, Bradley-Terry model describes the log-likelihood of the scale parameter $\pi = [\pi_1, ..., \pi_k], k \in [1, K]$ as

$$L(\pi) = \sum_i^K \sum_j^K [w_{ij} \log \pi_i - w_{ij} \log(\pi_i + \pi_j)] \tag{10}$$

[8] proposed to use MM-algorithm to find a maximum likelihood estimation of $\pi$ by performing an iterative update until convergence:

$$\pi_i^{(n+1)} = \frac{\sum_{j=1}^{K} w_{ij}}{\sum_{j=1}^{K} \frac{w_{ij}+w_{ji}}{\pi_i^{(n)}+\pi_j^{(n)}}} \tag{11}$$

Fitting a Bradley-Terry model with the results of the user study in Table 2 using (11), we can get an estimated preference scale of $\pi = (0.050, 0.445, 0.505)$ for L, SL and UL respectively. This result shows that the experts consider the images recovered from our proposed demosaicking method to have similar quality as the images from a supervised model, and the baseline linear demosaicking is the least favourable method.



\end{document}